\documentclass{article}
\begin{document}
\title{Dynamics of a Bose-Einstein  \\
              condensate in optical trap}
\author{F.Kh. Abdullaev $\dag$, B.B. Baizakov$\dag$, and V.V. Konotop$\ddag$ 
\\ $\dag$Physical-Technical Institute, Uzbek Academy of Sciences, \\ 2-b, G. Mavlyanov str., 700084 Tashkent, Uzbekistan 
\\ $\ddag$Departamento de F\'{\i}sica and Centro de F\'{\i}sica da Mat\'eria Condensada, \\
Universidade de Lisboa, Complexo Interdisciplinar,  \\ Avenida Profesor Gama Pinto, 2,
P-1649-003, Lisbon,Portugal}


\maketitle

\begin{abstract}
The dynamics of a 2D Bose-Einstein condensate in optical trap is studied taking
into consideration fluctuations of the far-off-resonance laser field intensity.
The problem is described in the frame of the mean field Gross-Pitaevskii
equation with randomly varying trap potential. An analytic approach based on
the moments method has been employed to describe the noise induced evolution
of the condensate properties. Stochastic parametric resonance
in oscillations of the condensate width is proved to exist. For the condensate
with negative scattering length of atoms, it is shown that the noise can
delay or even arrest the collapse. Analytical predictions are confirmed by
numerical simulations of the underlying PDE and ODE models.
\end{abstract}

\section{Introduction}
\label{Intro}

Recently a Bose-Einstein condensate has been realized in all-optical
far-off-resonance laser trap \cite{Stamper-Kurn98}. This important achievement
opens up new perspectives in exploring the phenomenon. On the one hand, with
the optical trap it is now possible to create more dense condensates, which is
valuable for investigation of three body decay processes, to obtain different
geometrical configurations including quasi-1D and 2D structures. On the other
hand, in an optical trap atoms in arbitrary hyperfine states may be confined,
therefore magnetic properties of atoms can be studied.

However, alongside with the above significant advantages, optical traps have
also some drawbacks, the most pertinent of which is related to fluctuations of
the laser field intensity. The last circumstance introduces stochasticity into
dynamical behavior of the condensate, which has to be taken into account in
real situations.

In the mean field theory of BEC based on the Gross-Pitaevskii (GP) equation,
fluctuations of the laser field intensity can be regarded as modulations of
the harmonic trap potential. The dynamics of BEC under regular variations of
trap parameters has been considered by many authors (see e.g. \cite{Castin96}
and references therein). From the experimental viewpoint the trap parameters
are modulated in order to reveal the response spectrum of the condensate.
Kagan {\em et al.,} \cite{Kagan96} have developed the scaling transformation
of the order parameter, which in 2D leads to easily solvable ODE's giving
results very close to exact numerical solution of the GP equation. The
existence of parametric resonances in the condensate response to time dependent
perturbations, on the ground of variational methods and numerical simulations,
was reported in Ref. \cite{Garcia-Ripoll99}.

In the present paper we study the dynamics of a BEC under randomly varying
optical trap potential, caused by fluctuations of the laser field intensity.
The basic idea consists in employing of differential relations
between several integral quantities of the GP equation. This approach,
known as {\em moments method}, will be applied to analysis of the evolution of
the atomic cloud's width driven by fluctuations of the laser field
intensity. The issues of particular interest include the possibility of the
stochastic parametric resonance in oscillations of the condensate width, and
whether the noise induced expansion of the atomic cloud can delay or
arrest the collapse of the BEC with attractive interaction between atoms.

\section{Basic equations and the Method of Analysis}
\label{Model}

Starting point to further development will be the fact, that the dynamics of
a nonlinear evolution equations (i.e. systems with infinite number of
degrees of freedom) can be exactly reduced to the dynamics of a system with
finite degrees of freedom \cite{Zakharov72}. Namely, by introducing collective
coordinates. This fact has been explored in Ref. \cite{Victor} in order to
illustrate the existence of extended parametric resonances in a two-dimensional
GP equation. An advantage of the model proposed in \cite{Victor} is that up to
certain extent it allows exact solutions.

We will concentrate on radially symmetric solutions ($\psi\equiv\psi(r,t)$)
to 2D GP equation, which describes the BEC having a cigar shape
\begin{equation} \label{main}
  i\frac{\partial \psi}{\partial t}=-\frac{1}{r}\frac{\partial}{\partial r}
  r\frac{\partial\psi}{\partial r}+(1+\epsilon(t))r^2\psi+\chi
  |\psi|^2\psi+i\gamma\psi .
\end{equation}

The dimensionless variables of this equation are related to actual quantities
through the following transformations: $r \rightarrow r/a_{ho}, \
t \rightarrow t\omega/2, \ \psi \rightarrow (8\pi N |a|/a_{ho})^{1/2} \psi$,
where $a_{ho} = \sqrt{\hbar /m\omega}$ is the harmonic oscillator unit, $\omega$
is the trap frequency, $a$ is the $s$-wave scattering length of atoms which
may be either positive or negative, $\epsilon(t)$ is the random function on
time simulating fluctuations of the laser trap frequency, $\chi=\pm 1$ carries
the sign of interaction type: "$+$" corresponds to a repulsive interaction
($a>0$) and "$-$" to attractive interaction ($a<0$), $\gamma$ is the
coefficient characterizing the damping effects due to influence of
non-condensed atoms (thermal cloud).

Modulation function $\epsilon(t)$ can be represented through the
deviations of the laser field intensity $E(t)$ around its mean
value $E_0$
\begin{equation}
\epsilon(t)=\frac{E(t)-E_0}{E_0} .
\end{equation}
In order to simplify the model, in what follows we concentrate on the case of
delta correlated random process with zero mean value:
\begin{equation} \label{sat}
  \langle \epsilon(t) \rangle = 0, \qquad
  \langle \epsilon(t) \epsilon(t^{\prime}) \rangle =
  2 \sigma^2 \delta(t-t^{\prime}),
\end{equation}
where $\sigma>0$ is the dispersion of fluctuations, $\delta(t)$ is
the Dirac delta function, and angular brackets stand for statistical
average over all realizations of the random process $\epsilon(t)$.
Then the following quantities are adequate to describe the evolution of the
Bose condensed atomic cloud
\begin{eqnarray} \label{In}
  I_0 &=& 2\pi\int_0^{\infty} |\psi|^2\,r\,dr , \\
  I_1 &=& 2\pi i\int_0^{\infty}(\psi\bar{\psi}_r-\bar{\psi}\psi_r)\, r^2\,dr,\\
  I_2 &=& 2\pi\int_0^{\infty}\left(|\psi_r|^2+
           \frac{\chi}{2}|\psi|^4 \right)r\,dr ,  \\
  I_3 &=& 2\pi\int_0^{\infty}|\psi|^2 r^3\,dr ,    \\
  I_4 &=& 2\pi\chi\int_0^{\infty}|\psi|^4 r\,dr .
\end{eqnarray}
These quantities correspond to the total number of atoms in the condensate,
radial field momentum, energy of the wave packet, mean square width of the
atomic cloud and mean field interaction energy respectively. They are governed
by the system of equations (hereafter a dot stands for the derivative with
respect to time)
\begin{eqnarray}
  \dot{I}_1 &=& -2\gamma I_1+4I_2-4(1+\epsilon(t))I_3 , \nonumber \\
  \dot{I}_2 &=& -2\gamma I_2-2(1+\epsilon(t))I_1-\gamma I_4 ,  \label{eqn} \\
  \dot{I}_3 &=& -2\gamma I_3+2I_1 , \nonumber
\end{eqnarray}
and the equation for $I_0$ is singled out
$$  \dot{I}_0=-2\gamma I_0 . $$
The above system is not closed in the presence of the
dissipation term, which will be considered below. We start with the
nondissipative case, $\gamma=0$ and derive the closed system for
averaged quantities $J_k=\langle I_k\rangle$ using for this
purpose the Furutsu-Novikov formula \cite{Klyatzkin80} which gives
\[ \langle\epsilon(t)I_k(t)\rangle=-4\sigma^2\delta_{k,3}
\langle I_k(t)\rangle
\]
($\delta_{k,l}$ is the Kronecker delta). Then the closed linear
system of equations for the averaged values reads
\begin{eqnarray} \label{sys}
  \dot{J}_1 =  4J_2 - 4J_3,           \quad
  \dot{J}_2 = -2J_1 + 8\sigma^2 J_3,  \quad
  \dot{J}_3 =  2J_1 .
\end{eqnarray}
This system is trivially solved. For the sake of definiteness we concentrate
on the quantity $J_3$ which in the case at hand can be interpreted as a mean
square width of the wave packet. Also, we consider the case of small noise
intensity $\sigma^2$, consistent with the real experimental situations,
then the solution has the form
\begin{equation} \label{J3}
  J_3(t)=Ce^{2\xi t}+C_1e^{-\xi t}\cos(\eta t)+C_2e^{-\xi t}\sin(\eta t),
\end{equation}
where $\xi$ and $\eta$ are introduced as a real and imaginary
part of the roots of the equation for eigenvalues
\begin{equation} \label{lambda}
  \lambda^3+16\lambda-64\sigma^2=0
\end{equation}
($\lambda_1=-2\xi$, $\lambda_2=\xi+i\eta$, and
$\lambda_3=\xi-i\eta$) associated with the linear system
(\ref{sys}). In particular for small dispersion $\sigma^2\ll 1$ we have
$\xi=2\sigma^2+O(\sigma^4)$, $\eta=4+O(\sigma^2)$.

Constants of integration $C_i$ are determined from initial conditions.
\begin{equation}
  J_3(0)=x_0^2, \qquad \dot J_3(0)=0, \qquad \ddot J_3(0) = 8J_2(0)-8J_3(0).
\end{equation}
If a Gaussian ansatz is assumed for the initial shape of the BEC
\begin{equation}
  \psi(r,t)=A(t)\exp\Bigl(-\frac{r^2}{2x(t)}\Bigr),
\end{equation}
where, the amplitude $A$ and width $x$ are connected through normalization
condition for the $\psi$ function, we have the following values for constants
\begin{eqnarray}
  C   \simeq x_0^2-\frac{2}{\eta^2}\bigl(x_0^2-\frac{1}{x_0^2}\bigr),
                                                               \qquad \nonumber
  C_1 \simeq \frac{2}{\eta^2}\bigl(x_0^2-\frac{1}{x_0^2}\bigr),\quad \nonumber
  C_2 \simeq -2x_0^2\frac{\xi}{\eta}. \nonumber
\end{eqnarray}
Then it follows from (\ref{J3}), that the noise enhances the effects in
the BEC. Noise induced expansion and contraction of the condensate
are described by the same eq.(\ref{J3}), "+" or "-" being assigned
to $\xi$, correspondingly. These conform with positive or negative
value of the energy $I_2$, therefore the sign in front of the last (noise)
term in eq.(\ref{lambda}) will be different in these two cases. The noise results
also in doubling of the frequency of oscillations of the condensate.

\section{Numerical modelling}
\label{Numerics}

In order to verify predictions of the analytical approach we performed direct
numerical solution of the GP equation (\ref{main}), as well as stochastic
ODE's (\ref{eqn}), resulting from the moments method. Due to increasing role
of effective procedures for integration of nonlinear and stochastic PDE's, we
briefly outline the numerical approach which has been used.

To proceed with evolution problems at first we find the ground state
solution to GP equation (\ref{main}) with stationary trap potential
($\epsilon(t) = 0$) in the absence of the damping term ($\gamma = 0$).
The corresponding ODE has the form
\begin{equation}
  \frac{d^2\psi}{dr^2} + \frac{1}{r}\frac{d\psi}{dr} - r^2\psi -
  \mid\psi\mid^2\psi - \beta = 0,
\end{equation}
where $\beta = 2\mu/\hbar \omega$ is the normalized chemical potential of the
condensate. In order to solve this equation the following boundary conditions
are applied: $\psi(0) = const, \ \psi^{'}(0)=0$, which state the regularity
and continuity of the $\psi$ - function at $r=0$. Taking into account the
fact, that asymptotic behavior of the $\psi$ - function follows that of the
2D harmonic oscillator solution, one can find the value for $\psi(0)$ as
described in Ref. \cite{Gammal99}.
To this end we applied the nonlinear equation solving procedure ZEROIN from
Ref.\cite{Forsythe77}, which combines advantages of the bisection and secant
techniques. After the both $\psi(0)$ and $\psi^{'}(0)$ are known, it is
strightforward to solve the equation using, for example Runge-Kutta procedure.
It is worth to mention, that this approach reproduces results for the
ground state wave functions presented in Refs. \cite{Gammal99,Ruprecht95}.

The obtained ground state solution has been used as an initial condition for
modelling of the time dependent GP equation (\ref{main}) with randomly
varying trap potential. In order to solve this PDE we employed the method of
lines, in which the equation is discretized along the spatial variable. From
400 to 800 grid points are taken to discretize the spatial variable within
the interval $r \in [0..5]$. Stability of the ground state solution is
checked by propagating eq.(\ref{main}) with $\epsilon(t)=0, \ \gamma=0$
up to $t=50$ in dimensionless time units. No detectable variation of the
ground state solution is observed. The resulting set of ODE's has been solved
by the procedure DOPRI8 \cite{Hairer87}, which is based on the Runge-Kutta
method of 6(7)th order. This procedure with adaptive step-size control is
proved to be efficient when providing the given accuracy in a less processor
time is a crucial requirement, which is the case in solving of stochastic
equations.

The system of equations (\ref{eqn}) when the damping term is neglected, can
be reduced to the singular Hill equation by introducing the variable
$X=\sqrt{I_3}$:
\begin{equation} \label{Hill}
  \ddot X + 4 (1+\epsilon(t)) X + Q/X^3 = 0,
\end{equation}
where $Q=2 I_2 I_3 - I_1^2/2$ is a constant, determined by integral
quantities or the initial data. The solution of this equation can be
represented in analytical form only for particular types of the modulation
function $\epsilon(t)$. For a stationary trap potential ($\epsilon(t)=0$)
the solution is
\begin{equation}
X(t) = \sqrt{\cos^2(2t)+\frac{Q}{4} \sin^2(2t)} .
\end{equation}
In the case of randomly varying trap potential the equation (\ref{Hill})
has to be solved numerically. We solved it using the procedure DOPRIN
\cite{Hairer87}, which is based on the Runge-Kutta method of 7(8)th order.

In Fig.1 the shapes of the trap potential
\begin{equation} \label{pot}
  U(x)=2 X^2 - \frac{Q}{2X^2} ,
\end{equation}
corresponding to equation (\ref{Hill}) at $\epsilon(t) = 0$
is shown for cases of repulsive and attractive condensates. From this figure
it is apparent the behavior of the above two types of condensates. In the
"particle in a potential well" representation, departures of the particle
from the stationary point (the minimum of the potential well), correspond
to expansions and contractions of the atomic cloud with positive $a$,
though it remains confined within the trap. Falling of the particle into the
center of a singular potential corresponds to the rapid convergence of the
condensate's width to zero (collapse of the condensate with negative $a$).

Fig.2 illustrates expansion of the condensate with a repulsive interaction
($a>0$) due to random variations of the trap potential. The gain parameter
$\xi$ and oscillation frequency $\eta$ determined from this figure,
\begin{figure}[h]
\caption{The shape of the potential curve eq. (\ref{pot}) for attractive
$Q = -1$ (dashed line) and repulsive $Q = 1$ (solid line) condensates.}
\label{fig1}
\end{figure}
which is the result of numerical solution of eq.(\ref{Hill}), averaged over
400 realizations of random paths, are in close agreement with predictions of
the moments approach eq.(\ref{J3}). The important conclusion coming out of
this result is the existence of the stochastic parametric resonance in the
system. Indeed, exponential growth of the amplitude and doubling of the
oscillations frequency are clear signatures of the stochastic parametric
resonance.
\begin{figure}[h]
\caption{Expansion of the condensate with repulsive interaction caused by
random variations of the trap frequency. Solid line is according to
formula (\ref{J3}), dashed line is the numerical solution of eq.(\ref{Hill}).
$X(0) = 0.707, \ {\dot X}(0)= 0, \ \sigma = 0.1, \ Q = 1$. }
\label{fig2}
\end{figure}

The evolution of the condensate wave function under randomly varying trap
potential, obtained by direct numerical solution of the GP equation
(\ref{main}), corresponding to a particular random path, is shown in Fig.3.
\begin{figure}[t]
\begin{minipage}[t]{60mm}
\end{minipage}
%
%
\begin{minipage}[t]{60mm}
\end{minipage}
\caption{Evolution of the condensate's profile under the action of random
variations of trap frequency. a) Numerical solution of eq.(\ref{main})
with $\gamma=0, \ \chi=+1 $ for a particular set of random numbers.
b) Variations of the condensate width, corresponding to this
path. $\sigma = 0.5.$ }
\label{fig3}
\end{figure}
Exponential growth of the condensate's width supplemented by doubling of
the frequency of oscillations, is evident also from this figure.
\begin{figure}[h]
\caption{Contraction of a 2D condensate with negative $s$-wave scattering
length under different strengths $\sigma$ of the optical trap's
noise. Initial conditions: $X(0)=1, {\dot X}(0)=0, Q=-1$.}
\label{fig4}
\end{figure}

The time dependence of the mean square width of the condensate, experiencing
collapse under the action of randomly varying trap potential, is presented
in Fig.4. Calculations are based on numerical solution of eq.(\ref{Hill}),
averaged over the ensemble of 400 realizations of random paths. In this case
we have the interplay between two opposite effects: contraction of the
condensate due to collapse and its stochastic expansion due to noise. As
it is apparent from this figure, increasing of the noise intensity $\sigma^2$
leads to slowing down of the collapse. At some level of the noise intensity,
contraction of the atomic cloud may be balanced, i.e the collapse can be
arrested.

\section{Conclusion}
\label{Concl}

In conclusion, we have studied the dynamics of a Bose-Einstein condensate
confined in an optical trap with randomly varying trap potential. Applying
the moments method the stochastic parametric resonance in oscillations of
the condensate width has been revealed. It is also established, that the
noise induced expansion of the condensate with attractive interaction between
atoms, can delay or even arrest the collapse.

\vspace{1 true cm}

F.Kh.A and B.B.B are grateful to the US CRDF (Award ZM2-2095) for
partial financial support of their work. V.V.K acknowledges support
from FEDER and Program PRAXIS XXI, No Praxis/P/Fis/10279/1998.


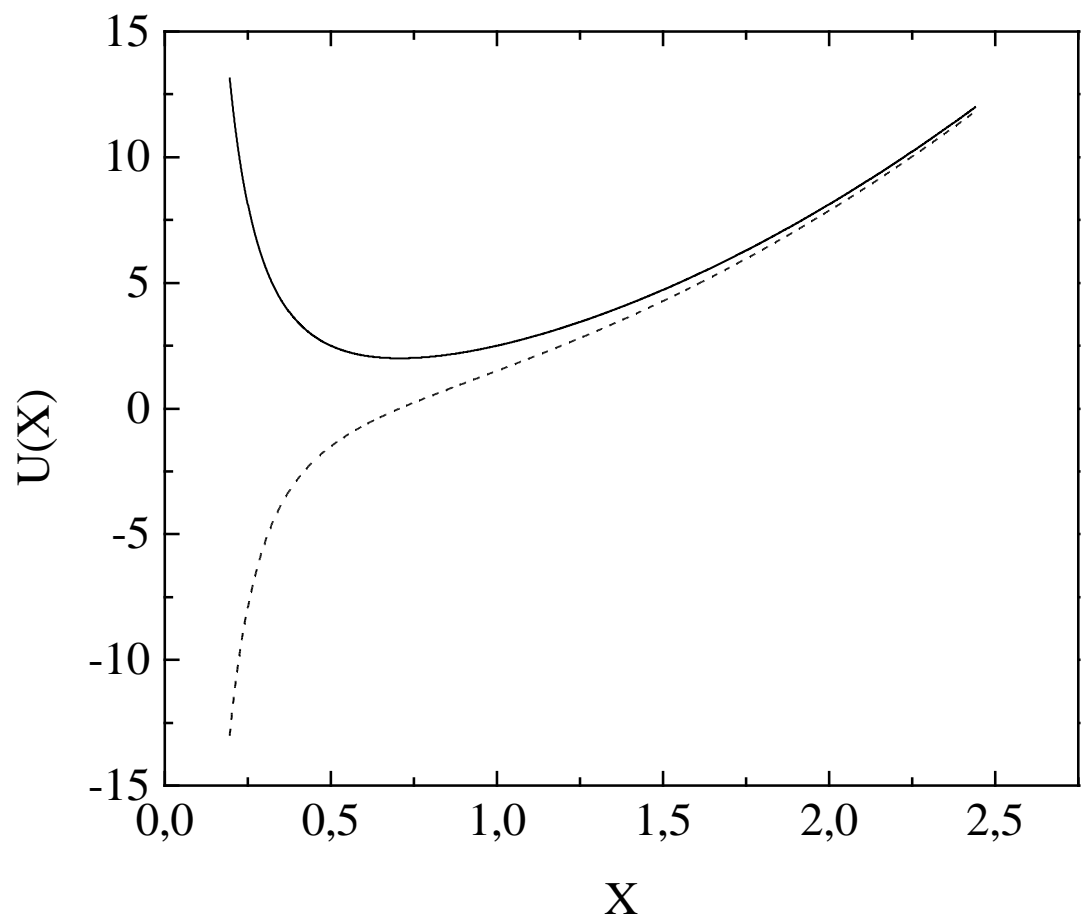
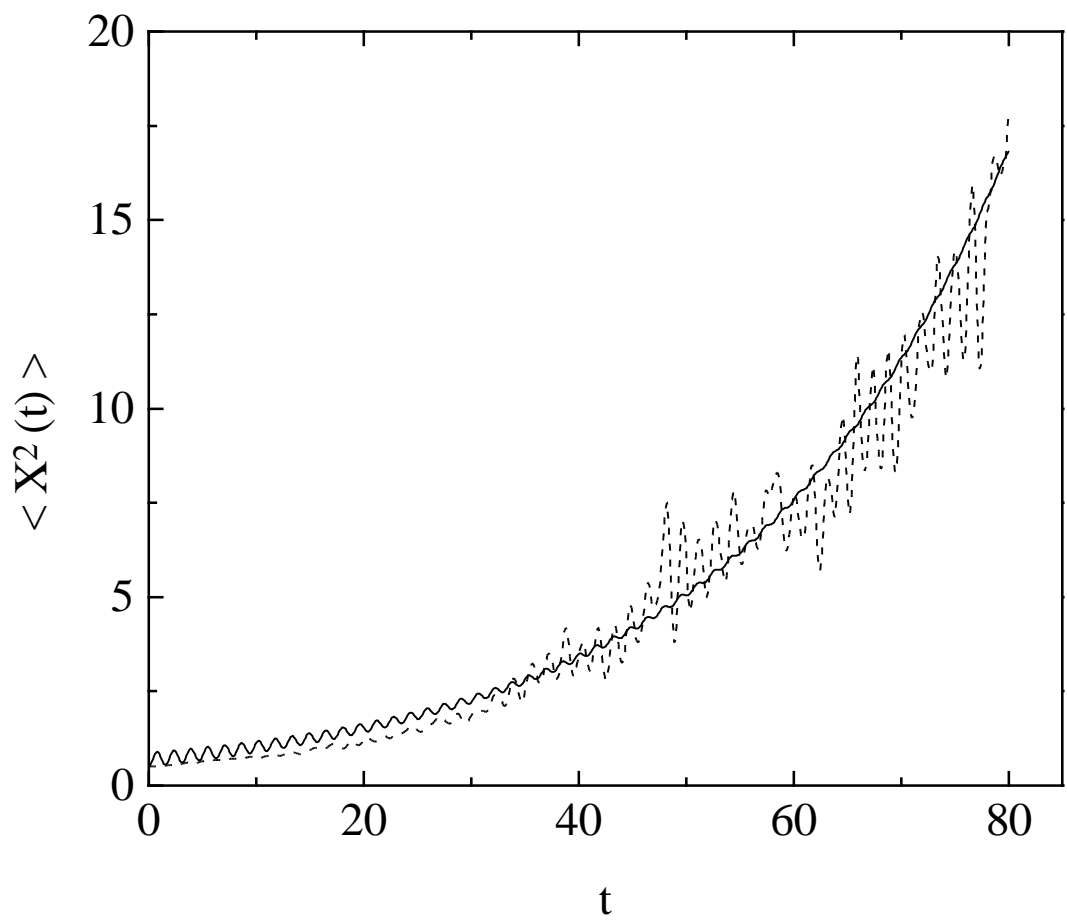
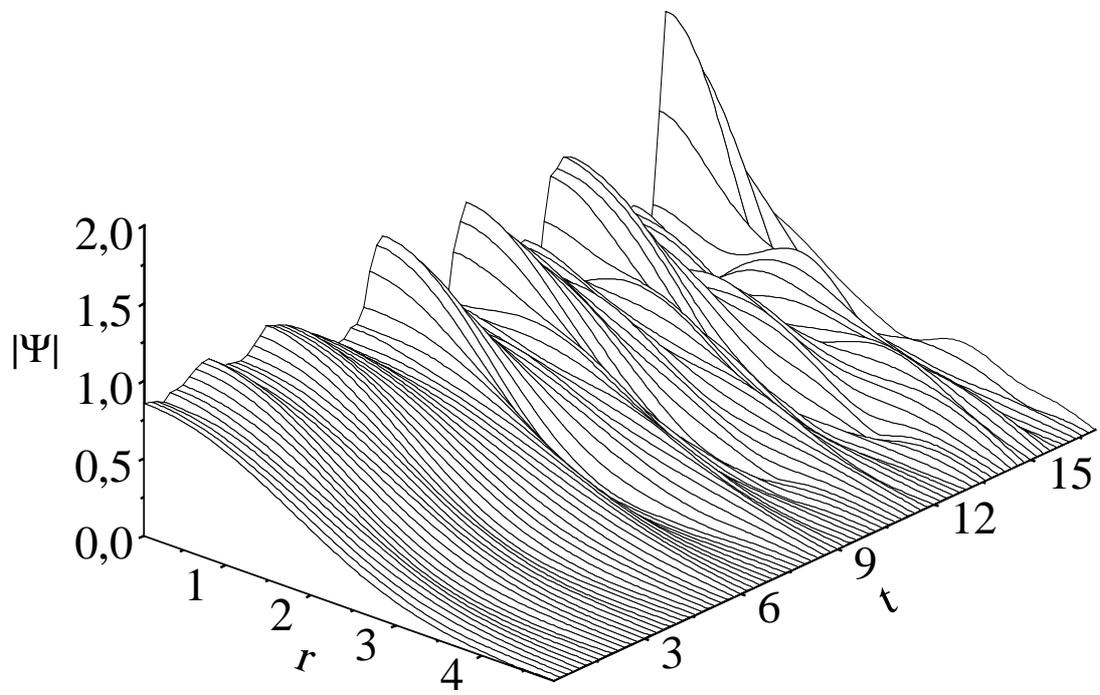
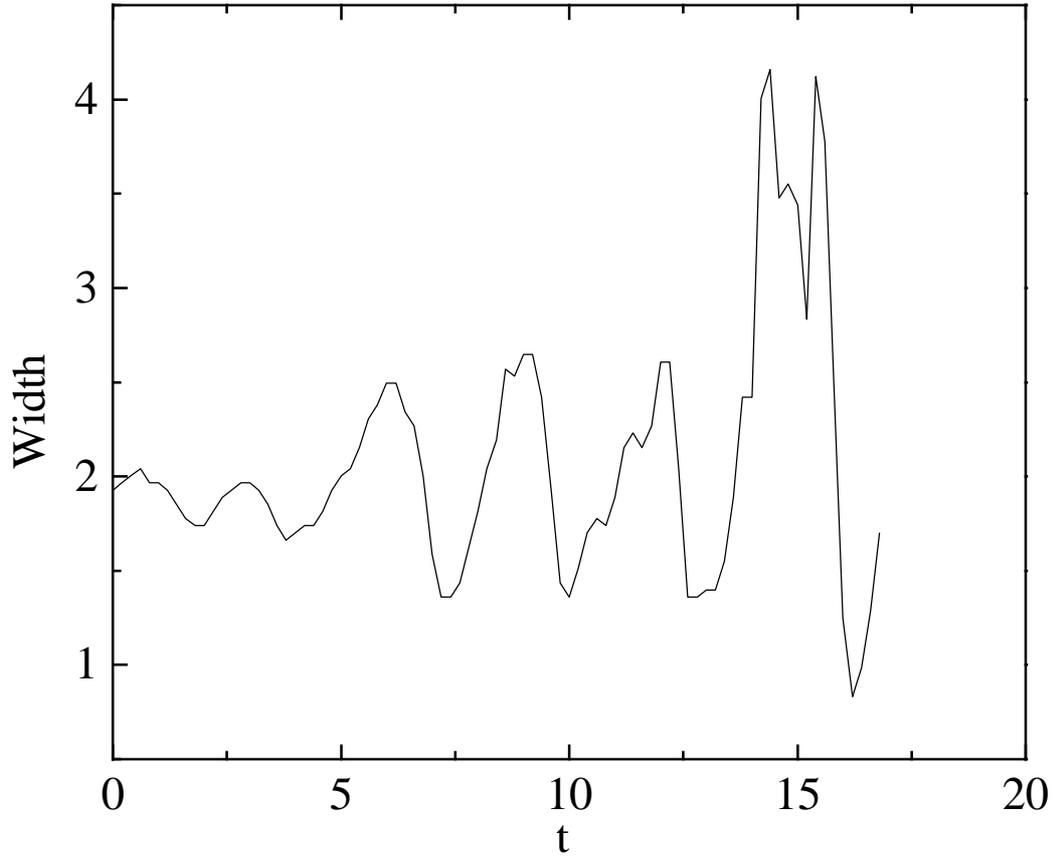
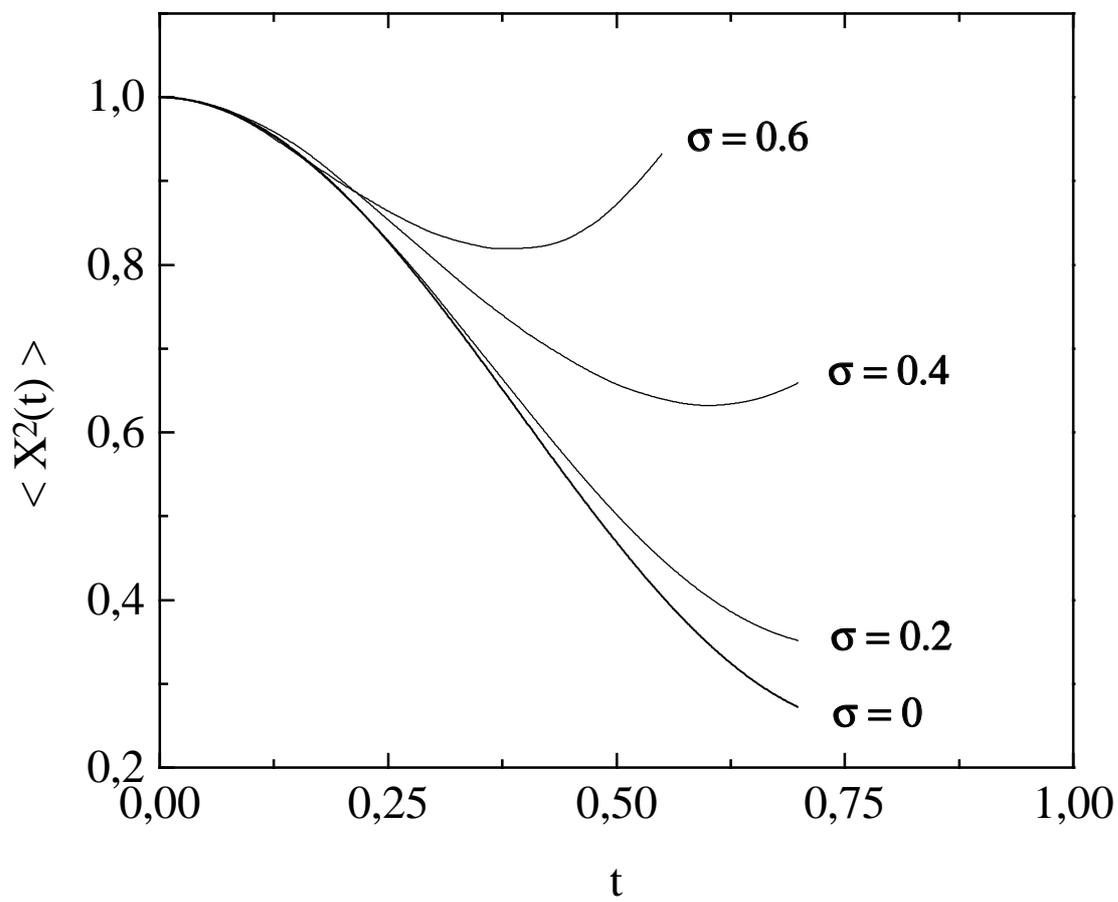
\end{document}